August 20, 2010

# Comment on "Perfect imaging with positive refraction in three dimensions"

R. Merlin

*Department of Physics, University of Michigan, Ann Arbor, MI 48109-1040*

Leonhardt and Philbin [Phys. Rev. A **81**, 011804(R) (2010)] have recently constructed a mathematical proof that the Maxwell's fish-eye lens provides perfect imaging of electromagnetic waves without negative refraction. In this comment, we argue that the unlimited resolution is an artifact of having introduced an unphysical drain at the position of the geometrical image. The correct solution gives focusing consistent with the standard diffraction limit.

August 20, 2010

In a recent paper, Leonhardt and Philbin (LP) claimed that the Maxwell's fish-eye lens [1] limited by a mirror behaves as a perfect lens for electromagnetic waves [2]. Their study extends an early report by Leonhardt for the two-dimensional case [3], and is consistent with work by Benítez et al. [4] who reached similar conclusions for scalar fields in various lensing systems. All these results appear to violate Abbe's diffraction limit [5] and to contradict Debye's classical theory of focusing [6,7]. Here, we show that the perfect focusing claimed in [2] is not an intrinsic property of the optical system, but the result of having added a point-dipole drain at the image position, which LP deem to be necessary to achieve a stationary state. In contrast, we find that the problem of a dipole in a mirror-bounded space does exhibit stationary states without the need for drains, and that the images obtained in that situation obey Abbe's constraint. Our analysis agrees with recent studies of focusing in metamaterials, which failed to observe deep subwavelength resolution in both the Maxwell's fish-eye and the Eaton lens [8].

Before examining the imaging properties of the mirror-coated fish eye itself, we consider the conceptually related but mathematically simpler problem of a point electric-dipole source at the center of a spherical cavity. We assume that the dipole is driven with a current whose time dependence is known a priori to be of the form $\exp(-i\omega t)$, and which cannot be modified by the radiation fields. The sphere, of radius $R$, is covered on the inside with a perfectly reflecting material. Ray optics dictates that the image is also at the center of the sphere. For a dipole of amplitude $p_0$ oriented along the $z$-axis, the solutions to the electromagnetic field equations are of the form $\mathbf{H} = H_\varphi e^{-i\omega t}\mathbf{e}_\varphi$ and $\mathbf{E} = \left(E_r \mathbf{e}_r + E_\theta \mathbf{e}_\theta\right)e^{-i\omega t}$ [9] with

$$H_\varphi = H_d + Ak^3 p_0 j_1(kr)\sin\theta$$
$$E_\theta = -\frac{i}{kr}\frac{\partial}{\partial r}(rH_\varphi)$$
$$E_r = i\frac{2H_\varphi}{kr\tan\theta}$$
(1)

Here, **H** and **E** are the magnetic and electric field, $k = \omega/c$ ($c$ is the speed of light), $j_1$ and $h_1^{(1)}$ are the first-order spherical Bessel and first-kind Hankel function, and $H_d = k^3 p_0 h_1^{(1)}(kr)\sin\theta$ is the contribution of the dipole alone which gives outgoing waves $\propto e^{i\omega(r/c-t)}/kr$ for $kr \gg 1$. From the boundary condition, requiring that $E_\theta = 0$ at $r = R$, we get

$$A = -\frac{h_1^{(1)}(kR) + kRh_1^{(1)'}(kR)}{j_1(kR) + kRj_1'(kR)}$$

where the prime indicates the derivative with respect to $kR$. LP assert generally that "to maintain a stationary regime, we must supplement the source by a drain" [2]. However, given that the fields in Eq. (1) are drain-free and that a solution exists, except for the cavity resonances at $j_1(kR) + kRj_1'(kR) = 0$, it is clear that the LP statement is incorrect. Hence, there is no need to include a drain to achieve a steady state [10].

Taking a path that parallels that of LP in their treatment of the fish-eye lens, we now add at the center of the sphere an electric-dipole drain (a time-reversed source) with amplitude identical to that of the original source. Then, the solution is

$$H_\varphi = k^3 p_0 \left[ h_1^{(1)}(kr) + e^{i\Phi} h_1^{(2)}(kr) \right]\sin\theta \ .$$
(2)

where

August 20, 2010

$$e^{i\Phi} = -\frac{h_1^{(1)}(kR) + kR h_1^{(1)\prime}(kR)}{h_1^{(2)}(kR) + kR h_1^{(2)\prime}(kR)} \ .$$

The first-order spherical Hankel function of the second kind, $h_1^{(2)}$, describes the incoming waves associated with the drain; for $kr \gg 1$, $h_1^{(2)} \sim e^{-i\omega(r/c+t)}/kr$. Similar to the LP results for the fish-eye, the phase shift is $\Phi \approx 2kR + \pi$ for $kR \gg 1$ so that, in this limit, the magnetic field vanishes at $r = R$.

In Eq. (1), the non-singular solution to the homogeneous equation, $j_1(kr)$, represents the field due to the induced currents at the inner surface of the sphere. It is apparent that this term gives a diffraction-limited image at $r = 0$, whereas the presence of the drain in Eq. (2) leads to the disappearance of the image field and, for $kR \gg 1$, of the induced currents.

Since any linear combination of $h_1^{(1)}$, $h_1^{(2)}$ and $j_1$ can be written as a sum involving only two of these functions, the question arises as to which one of the infinite possible partitions among source, drain and image is the correct description of a particular problem. Some reflection shows that the association of the incoming wave to a drain in Eq. (2), something we have done following LP [2,3], is but an inaccurate assessment of what is actually a consequence of the insertion of a second electric dipole at $\mathbf{r} = 0$. The amplitude and phase of this new source are such that, when the field it produces is added to that of the induced currents, the resulting field mimics the behavior of a time-reversed source [11]. To see this, we note that the fields of Eq. (1) satisfy the wave equation for the Hertz vector $\Pi$ with a single source term

$$\nabla^2 \Pi^{(1)} - \frac{1}{c^2}\frac{\partial^2}{\partial t^2}\Pi^{(1)} = -4\pi\delta(\mathbf{r}) p_0 e^{i\omega t}\mathbf{e}_z$$

while those of Eq. (2) are the solutions of

$$\nabla^2 \Pi^{(2)} - \frac{1}{c^2} \frac{\partial^2}{\partial t^2} \Pi^{(2)} = -4\pi\delta(\mathbf{r})\left(p_0 - p_0 e^{i\Phi}\right) e^{i\omega t + \gamma t} \mathbf{e}_z \quad,$$

which contains a second source that emits radiation delayed by $\Phi$ with respect to the original one (the two dipoles coincide here because both object and image are at $\mathbf{r} = 0$; in the fish-eye, the conjugate foci and, thus, the dipoles occupy different positions [2,3]). If we combine the field due to the second source alone with that of the currents induced by the pair of dipoles, we get

$$H_\varphi = k^3 p_0 \left[ A\left(1 - e^{i\Phi}\right) j_1(kr) - e^{i\Phi} h_1^{(1)}(kr) \right] \sin\theta = k^3 p_0 e^{i\Phi} h_1^{(2)}(kr) \sin\theta \quad.$$

This contribution, representing the field of a source operating in reverse, is what LP call a drain.

It is then clear that Eq. (1) and Eq. (2) are physical, causal solutions to different problems, none of which contains a drain. Thus, interpretations of imagining involving drains are problematic because they either violate causality (an electric dipole that did not exist in the past appears out of thin air in the future to provide perfect imaging) or require the introduction of radiation sources other than the object [11]. We note that conventional absorbers, as considered in [2-4], are not drains. An absorbing subwavelength particle acquires electric and magnetic dipole moments proportional to the respective fields (which can be considered uniform), but with a phase delay in the range $\pi < \varphi < 2\pi$. As such, small absorbers behave both as energy sinks and radiation sources. Hence, for imaging purposes, the only physically sensible depiction of the various field components is that given by Eq. (1).

The solution with equal source and drain strength, Eq. (2), is analogous to what LP found for the mirror-coated fish eye. It is easy to see that what LP refer to as "perfect imaging" is nothing more than the singularity introduced by the drain or, better, by the unjustified addition of an extra source of radiation. Let $\mathbf{H} = \mathbf{H}_{\text{LP}} \exp(-i\omega t)$ be the LP solution for the magnetic field for which the source and drain are at the position of the object ($\mathbf{r}_O$) and the image ($\mathbf{r}_I$), respectively. Since the solution that exchanges the source and drain, $\mathbf{H}_{LP}^{*}\exp(-i\omega t)$, also satisfies Maxwell's equations, it is clear that we can construct linear combinations of these two solutions that do not have a singularity at the image position. Specifically, if $\mathbf{H}_{LP}$ behaves as $H_{\varphi_O} \approx k^3 p_0 h_1^{(1)}(k\tilde{r}_O)\sin\theta_O$ and $H_{\varphi_I} \approx k^3 p_0 e^{i\Psi} h_1^{(2)}(k\tilde{r}_I)\sin\theta_I$ in the vicinity of $\mathbf{r}_O$ and $\mathbf{r}_I$, then the desired combination is $i\left(e^{-i\Psi}\mathbf{H}_{LP} + e^{+i\Psi}\mathbf{H}_{LP}^{*}\right)/2\sin\Psi$ giving

$$H_{\varphi_O} \approx k^3 p_0 \left[ h_1^{(1)}(k\tilde{r}_O) + \frac{ie^{i\Psi}}{\sin\Psi} j_1(k\tilde{r}_O) \right] \sin\theta_O$$
$$H_{\varphi_I} \approx \frac{ik^3 p_0}{\sin\Psi} j_1(k\tilde{r}_I)\sin\theta_I \quad . \tag{3}$$

Here, the spherical coordinates are defined with respect to the orientation of the corresponding dipoles, $\Psi$ is a phase delay [2] and $\tilde{r}_{O,I} = |\mathbf{r} - \mathbf{r}_{O,I}|$. These drain-free expressions describe a source at $\mathbf{r}_O$ and two diffraction-limited images at $\mathbf{r}_O$ and $\mathbf{r}_I$. It follows from the spherical-cavity discussion that this (and not the LP) solution is the physically correct answer to the fish-eye lens problem.

*Note added.* After this paper was submitted, the author became aware of work by Blaikie [12] questioning Leonhardt's claim of perfect imaging for the two-dimensional version of Maxwell's fish-eye [3]. For Leonhardt's Reply to Blaikie's Comment, see [13].

**Acknowledgments.** Work supported by the Air Force Office of Scientific Research under contract FA 9550-06-01-0279 through the Multidisciplinary University Research Initiative Program.